\documentclass[%
 reprint,
superscriptaddress,
 amsmath,amssymb,
 aps,
 prapplied,
 longbibliography,
 floatfix,
]{revtex4-2}

\usepackage{graphicx}% Include figure files
\usepackage{dcolumn}% Align table columns on decimal point
\usepackage{bm}% bold math
\usepackage{hyperref}% add hypertext capabilities
\usepackage{braket}
\usepackage{physics}
\usepackage{amsmath}
\renewcommand{\vec}[1]{\boldsymbol{#1}}
\newcommand{\dqc}{Duke Quantum Center, Duke University, Durham, North Carolina 27701, USA}

\begin{document}

\preprint{APS/123-QED}

\title{Batch Optimization of Frequency-Modulated Pulses for Robust Two-qubit Gates in Ion Chains}

\author{Mingyu Kang}
\email{mingyu.kang@duke.edu}
\affiliation{\dqc}
\affiliation{Department of Physics, Duke University, Durham, North Carolina 27708, USA}
\author{Qiyao Liang}%
\affiliation{\dqc}
\affiliation{Department of Physics, Duke University, Durham, North Carolina 27708, USA}
\author{Bichen Zhang}%
\email{bichen.zhang@duke.edu}
\affiliation{\dqc}
\affiliation{Department of Electrical and Computer Engineering, Duke University, Durham, North Carolina 27708, USA}
\author{Shilin Huang}%
\affiliation{\dqc}
\affiliation{Department of Electrical and Computer Engineering, Duke University, Durham, North Carolina 27708, USA}
\author{Ye Wang}
\affiliation{\dqc}
\affiliation{Department of Electrical and Computer Engineering, Duke University, Durham, North Carolina 27708, USA}
\author{Chao Fang}
\affiliation{\dqc}
\affiliation{Department of Electrical and Computer Engineering, Duke University, Durham, North Carolina 27708, USA}
\author{Jungsang Kim}
\affiliation{\dqc}
\affiliation{Department of Physics, Duke University, Durham, North Carolina 27708, USA}
\affiliation{Department of Electrical and Computer Engineering, Duke University, Durham, North Carolina 27708, USA}
\affiliation{IonQ, Inc., College Park, Maryland 20740, USA}
\author{Kenneth R. Brown}%
\email{ken.brown@duke.edu} 
\affiliation{\dqc}
\affiliation{Department of Physics, Duke University, Durham, North Carolina 27708, USA}
\affiliation{Department of Electrical and Computer Engineering, Duke University, Durham, North Carolina 27708, USA}
\affiliation{Department of Chemistry, Duke University, Durham, North Carolina 27708, USA}

\date{\today}

\begin{abstract}

Two-qubit gates in trapped-ion quantum computers are generated by applying spin-dependent forces that temporarily entangle the internal state of the ion with its motion. Laser pulses are carefully designed to generate a maximally entangling gate between the ions while minimizing any residual entanglement between the motion and the ion. The quality of the gates suffers when the actual experimental parameters differ from the ideal case. Here, we improve the robustness of frequency-modulated Mølmer-Sørensen gates to motional mode-frequency offsets by optimizing the average performance over a range of systematic errors using batch optimization. We then compare this method with frequency-modulated gates optimized for ideal parameters that include an analytic robustness condition. Numerical simulations show good performance up to 12 ions, and the method is experimentally demonstrated on a two-ion chain.
\end{abstract}

\maketitle

\section{Introduction}
Trapped-ion systems are one of the leading candidates for a scalable quantum computing platform \cite{Monroe13, Brown16}. In addition to near-perfect coherence properties \cite{Wang17, Wang21} and single-qubit gates with error rates below $10^{-4}$ \cite{Brown11, Harty14, Mount15, AudeCraik17}, trapped-ion qubits have significant advantages in terms of entangling-gate fidelities. For systems with exactly two ions, state-of-the-art two-qubit gates have reached a fidelity higher than 99.9\% when a state-dependent force is applied with lasers \cite{Gaebler16, Ballance16} or magnetic field gradients \cite{Srinivas21}. For larger systems, two-qubit gate fidelities greater than 99\% for a four-ion chain \cite{Wang20} and greater than 97\% for 13-ion and 17-ion chains \cite{Wright19, Landsman19} have been reported. Trapped-ion systems with many qubits are particularly promising, as long-range Coulomb interactions between ions lead to all-to-all connectivity between qubits \cite{Wright19, Linke3305}.

The central challenge in achieving scalability is to perform high-fidelity entangling gates with a large number of qubits. Entangling gates are performed by briefly exciting the ions' normal modes of motion, which serve as a carrier of quantum information \cite{Molmer99, Sorensen99}. The driving field needs to be carefully controlled such that all motional modes are completely disentangled from the internal qubit states at the end of the gate, while the qubit states undergo a maximally entangling operation. 

In the presence of noise and parameter drifts, pulse design is necessary to achieve fast and robust high-fidelity gates. One approach is to design the amplitudes of multichromatic beams that suppress the effect of noise \cite{Haddadfarshi16, Webb18, Shapira18, Zarantonello19, Blumel19, Shapira20}. Another way is to control the amplitude \cite{Zhu06, Roos08, Kim09, Choi14, Debnath16, Figgatt19}, phase \cite{Green15, Lu19, Milne20-P, Bentley20}, and/or frequency \cite{Wang20, Leung18-R, Landsman19} modulation over many time segments; this has recently been applied in experiments with many ions \cite{Choi14, Debnath16, Figgatt19, Landsman19, Lu19, Wang20}. 

While the methods above lead to analytic robustness by guaranteeing high fidelity up to a certain order \cite{Blumel19} for the uncertainty in a control parameter, a promising approach is to find a robust pulse numerically using optimization algorithms inspired by machine learning (ML). In particular, Ref. \cite{Wu-Ding19} showed that training with a large sample set and minibatches of parameter offsets significantly improves the robustness of the optimized pulse for a generic Hamiltonian with control fields. For trapped-ion systems, Ref. \cite{Ai21} demonstrated the application of deep reinforcement learning to a robust single-qubit gate. 

In this paper, we improve on previous discrete and continuous frequency-modulation (FM) schemes \cite{Wang20, Leung18-R}. We propose two algorithms for FM pulse optimization using training with a large sample set and with minibatches, namely, s(ample)-robust and b(atch)-robust FM, following the notation of Ref. \cite{Wu-Ding19}. The rest of the paper is organized as follows. In Section \ref{sec:theory}, we briefly review the theory of robust frequency-modulated Mølmer-Sørensen (MS) gates \cite{Leung18-R} and introduce the optimization schemes for s-robust and b-robust FM. In Section \ref{sec:opt}, we show that s-robust and b-robust FM are significantly more robust than the previous robust FM to motional mode frequency drifts. We also discuss the scalability of b-robust FM. In Section \ref{sec:experiment}, we show experimental results for a two-ion chain that demonstrate that b-robust FM is more robust than robust FM to detuning errors. Finally, we summarize our results and discuss future directions in Section \ref{sec:conclusion}.

\section{Robust optimization methods for frequency-modulated MS gates}\label{sec:theory}

The frequency-modulated MS gate uses a state-dependent force induced by lasers at a drive frequency, modulated near sideband frequencies \cite{Molmer99, Sorensen99}. When addressed by lasers with the correct optical phases, the ions $j_1$ and $j_2$ undergo the unitary evolution described by the following equation \cite{Leung18-R, Wu18}:

\begin{align}\label{eq:U}
    U(\tau) = \exp &\Big\{\sum_{j=j_1,j_2}\sum_k 
    \Big( [\alpha^j_k(\tau)\hat{a}_k^\dagger - \alpha^{j*}_k(\tau)\hat{a}_k] \:\hat{\sigma}^j_x\Big) \nonumber\\
    &\quad+ i\Theta(\tau)\:\hat{\sigma}^{j_1}_x\hat{\sigma}^{j_2}_x \Big\},
\end{align}
where
\begin{equation}\label{eq:alpha}
    \alpha^j_k(\tau) = \frac{\Omega}{2}\eta^j_k \int^\tau_0 e^{-i\theta_k(t)}dt,
\end{equation}

\begin{equation}\label{eq:Theta}
    \Theta(\tau) = -\frac{\Omega^2}{2}\sum_k \eta^{j_1}_k \eta^{j_2}_k \int^\tau_0 dt_1 \int^{t_1}_0 dt_2 \sin[\theta_k(t_1) - \theta_k(t_2)].
\end{equation}
Here, $\tau$ is the pulse length, $\Omega$ is the carrier Rabi frequency, $\eta^j_k$ is the Lamb-Dicke parameter of ion $j$ with respect to motional mode $k$, and $\hat{\sigma}^j_x$ is the bit-flip Pauli operator of ion $j$. Also, 
\begin{equation}\label{eq:theta}
    \theta_k(t) = \int^t_0 [\mu(t') - \omega_k] dt'
\end{equation}
is the phase of motional mode $k$, which is the integral of the detuning between the drive frequency $\mu(t)$ and the mode frequency $\omega_k$. The first term in Eq. \ref{eq:U} describes state-dependent displacement of the motional modes, while the second term represents rotation with respect to the two-qubit axis $\hat{\sigma}^{j_1}_x\hat{\sigma}^{j_2}_x$.

For an ideal MS gate, the qubits should be completely disentangled from the motional modes \mbox{[$\alpha^j_k(\tau) = 0 \: \forall j,k$]}, and the rotation angle $\Theta(\tau)$ should reach exactly $\pi/4$ at the gate's conclusion \cite{Figgatt19, Grzesiak20}. Hence, the goal of robust FM is to modulate the drive-frequency profile $\mu(t)$ such that $\alpha^j_k(\tau)$ and \mbox{$|\Theta(\tau)-\pi/4|$} are sufficiently minimized in the presence of mode-frequency offsets $\epsilon_k$, i.e., \mbox{$\omega_k \rightarrow \omega_k + \epsilon_k$}. 

Minimizing \mbox{$|\alpha^j_k(\tau)| \propto |\int^\tau_0 e^{-i \theta_k(t)}dt|$} is the intuitive criterion for an optimized gate. However, such a gate is sensitive to small changes $\epsilon_k \ll 1/\tau$. Instead, the authors of Ref. \cite{Leung18-R} induce robustness by minimizing the time-averaged displacement \mbox{$|\alpha^j_{k,\text{avg}}| \propto \frac{1}{\tau}|\int^\tau_0 \int^t_0 e^{-i\theta_k(t')}dt' dt|$}, which is proportional to the first-order correction to \mbox{$|\alpha_k^j(\tau)|$ when $\omega_k \rightarrow \omega_k + \epsilon_k$}. Note that a time-symmetric pulse can be used to guarantee that minimizing $|\alpha^j_{k,\text{avg}}|$ also minimizes $|\alpha^j_k(\tau)|$. This optimization scheme, which we call ``robust FM,'' has been used in recent experiments with four-ion \cite{Wang20} and 17-ion \cite{Landsman19} chains. Similar approaches with amplitude and phase modulation \cite{Bentley20, Milne20-P} have also been studied. 

Although robust FM has been shown to be robust to mode-frequency offsets that are an order of magnitude smaller than $1/\tau$, it does not guarantee robustness to $\epsilon_k \lesssim 1/\tau$. Moreover, robustness of the angle $\Theta(\tau) \approx \pi/4$ to detuning errors is not enforced by this method.

Inspired by recent work on applying machine learning with a large sample set and minibatches to quantum control \cite{Wu-Ding19}, we present ``s(ample)-robust'' and ``b(atch)-robust'' FM, which further enhance the robustness of a two-qubit gate. Instead of minimizing the analytic first-order correction, we minimize the average of $|\alpha^j_k(\tau)|^2$ over an ensemble of offsets, thereby directly incorporating the robustness condition into the cost function. Similarly, we also include the condition for robustness of the angle $\Theta(\tau)$ in our cost function. Note that optimizing robustness of the displacement has been achieved to some extent by various methods \cite{Shapira18, Blumel19, Milne20-P}, but not with the additional goal of optimizing robustness of the angle \footnote{In Ref. \cite{Shapira18} and supplementary information of Ref. \cite{Blumel19}, the residual entanglement, or the displacement error, is removed up to a certain order in the motional frequency offset. However, the rotation-angle error is not removed; only the pre-factor of the leading-order term in the motional frequency offset is minimized to a nonzero value.}, although this is crucial for reaching high fidelity in the presence of motional frequency drifts. We find the optimal FM pulse $\mu(t)$ that minimizes the following cost function $C_{\mathcal{E}}$:

\begin{align}\label{eq:C}
    &\quad\quad\quad\quad\quad\quad\quad\quad
    C_{\mathcal{E}} =  \frac{1}{|S_{\mathcal{E}}|} \sum_{\vec{\epsilon} \in S_{\mathcal{E}}} C(\vec{\epsilon}), \nonumber
    \\
    &C(\vec{\epsilon}) = 
    \sum_k \big(\alpha^{j_1}_k(\tau, \vec{\epsilon})^2 + \alpha^{j_2}_k(\tau, \vec{\epsilon})^2\big) +
    \frac{1}{2}\big(\Theta(\tau, \vec{\epsilon}) - \frac{\pi}{4}\big)^2 .
\end{align}

Here, $\mathcal{E}$ is the motional frequency uncertainty, and $S_\mathcal{E}$ consists of offset vectors $\vec{\epsilon}$ whose components $\epsilon_k$ are independently and randomly drawn from the normal distribution $\mathcal{N}(0, \mathcal{E})$. $\alpha^{j}_k(\tau, \vec{\epsilon})$ and $\Theta(\tau, \vec{\epsilon})$ are the displacement and angle when $\omega_k \rightarrow \omega_k + \epsilon_k$. The two terms of $C(\vec{\epsilon})$ are simply the displacement error representing residual entanglement with the phonons, and the angle error. 

The carrier Rabi frequency $\Omega$ is updated at each iteration such that $\Theta(\tau, \vec{0}) = \pi/4$. Since the displacement error is proportional to $\Omega^2$ and the angle error is proportional to $\Omega^4$, this cost function naturally finds the low-$\Omega$ solution. This differs from the robust FM approach, which sets $\Omega$ after the entire optimization \cite{Leung18-R}, requiring explicit regularization to fit the experimental constraints. 

For s-robust FM, we set $S_\mathcal{E}$ to a fixed training set throughout the optimization. For b-robust FM, we set $S_\mathcal{E}$ to a batch, which is randomly updated at each iteration of the optimization. Therefore, while s-robust FM calculates the cost function repeatedly with a certain set of samples, b-robust FM computes the cost function with a different batch generated from the error distribution throughout the entire optimization. In the work presented here, we set the training-set size to 100 for s-robust FM and the batch size to 10 for b-robust FM. For the batch method, we choose the adaptive-moment-estimation \cite{ADAM} optimizer to stabilize the gradient during training. We obtain sufficiently good results without hyperparameter tuning.

\section{Comparison of optimization methods} \label{sec:opt}

\begin{figure}[ht]
\includegraphics[width=8.6cm]{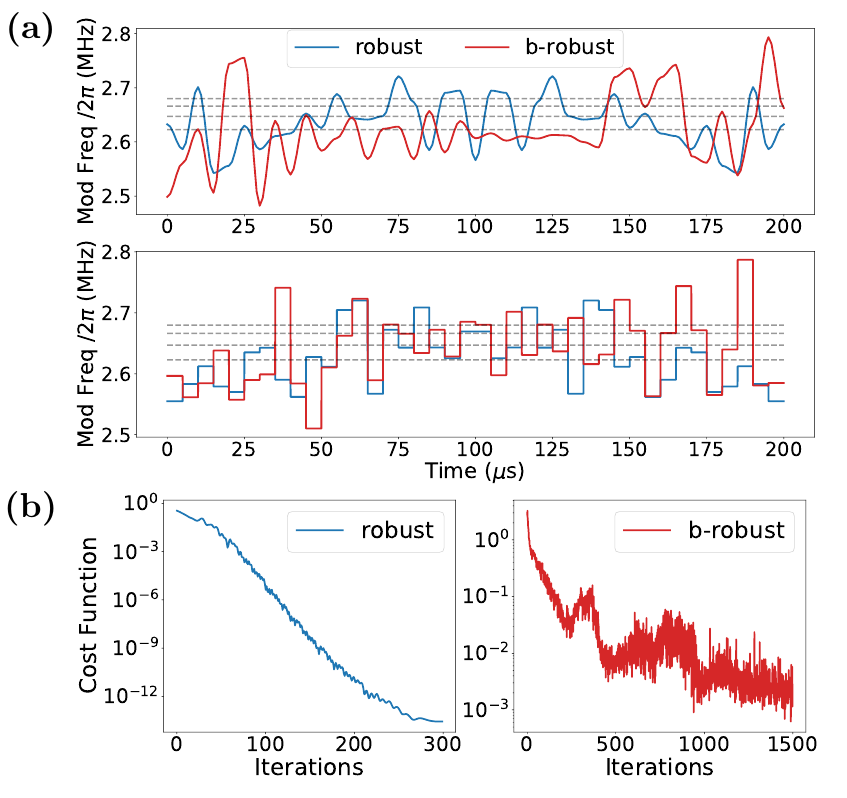}
\caption{(a) Continuous (top) and discrete (bottom) pulses optimized by robust and b-robust FM. The gray lines show the sideband frequencies for a four-ion chain. Robust FM pulses are time-symmetric, while b-robust FM pulses are not. (b) Learning curves for robust and b-robust FM optimization.}
\label{fig:1}
\end{figure}

Fig. \ref{fig:1}a shows examples of continuous and discrete pulses from robust and b-robust FM optimization over a four-ion chain. Note that b-robust pulses do not have the even-pulse constraint and thus have twice as many degrees of freedom as robust pulses. This allows b-robust FM to explore a wider range of pulse shapes. 

The continuous and discrete pulses have different time complexities for evaluating the gradient of the angle $\Theta(\tau)$ over the pulse $\mu(t)$, which is the most time-consuming routine of the optimization. For continuous pulses, neighboring steps are connected by substeps that follow a cosine envelope, and the evaluation time is quadratic in the number of substeps. However, for discrete pulses, the stepwise-constant form allows efficient evaluation of the gradient of $\Theta(\tau)$, requiring time linear in the number of steps. 

Fig. \ref{fig:1}b shows the learning curves for robust and b-robust optimization. For robust FM, the cost function quickly and smoothly drops to lower than $10^{-6}$. However, this guarantees a very accurate gate solution only at offsets close to zero. Meanwhile, for b-robust FM, we set the motional frequency uncertainty $\mathcal{E}$ as $2\pi \times 1$ kHz. The cost function experiences larger fluctuations, as a new batch of parameter offsets is used for optimization at each iteration. Although the cost function reaches only approximately $10^{-3}$, we expect our gate fidelity to be robust, \mbox{$\mathcal{F} \geq 1-10^{-3}$}, against all mode-frequency offsets within the optimized range. 

%%%%%%%%%%%%%%%%FIGURE 1 %%%%%%%%%%%%%%%%
%%%%%%%%%%%END OF FIGURE 1%%%%%%%%%%%%%%%%

To evaluate the robustness performance, we calculate the average unitary gate fidelity over the test set $T_{\mathcal{E}}$ of mode-frequency offsets. The unitary fidelity can be expressed as \mbox{$\mathcal{F} = \frac{1}{D}\Tr(\mathcal{U}^\dagger U(\tau))$}, where $U(\tau)$ is the unitary evolution in Eq. \ref{eq:U}, $\mathcal{U}$ is the target unitary, and $D$ is the Hilbert-space dimension \cite{Rabitz05}. In terms of displacement and angle, the average fidelity can be expressed to second order as in the following equation (see the supplementary information of Ref. \cite{Bentley20} for the derivation):
\begin{align}
     \mathcal{F}_\mathcal{E} = &\frac{1}{|T_{\mathcal{E}}|} \sum_{\vec{\epsilon} \in T_{\mathcal{E}}} \mathcal{F}(\vec{\epsilon}), \nonumber 
     \\
    \mathcal{F}(\vec{\epsilon}) = &\cos\Big(\Theta(\tau, \vec{\epsilon}) - \frac{\pi}{4}\Big) \nonumber
    \\
    &\times \Bigg[
    1 - \sum_k \Big(\alpha^{j_1}_k(\tau, \vec{\epsilon})^2 + \alpha^{j_2}_k(\tau, \vec{\epsilon})^2 \Big)\Big(\overline{n}_k + \frac{1}{2}\Big) \Bigg],
\end{align}
where $\overline{n}_k$ is the mean phonon number of mode $k$, and $T_\mathcal{E}$ is the test set of the motional frequency uncertainty $\mathcal{E}$, constructed similarly to $S_\mathcal{E}$. In order to evaluate the robustness, the test set is completely random and independent of the training set or minibatches used for optimization. We choose the test-set size to be 1000. For an initial state with an average of 0.5 phonons, the fidelity is simply equal to 1 minus the cost function to leading order in the errors: \mbox{$\mathcal{F}(\vec{\epsilon}) = 1-C(\vec{\epsilon})$}.

%%%%%%%%%%%%%%%%FIGURE 2 %%%%%%%%%%%%%%%%
\begin{figure*}[ht!]
\includegraphics[height=7.5cm]{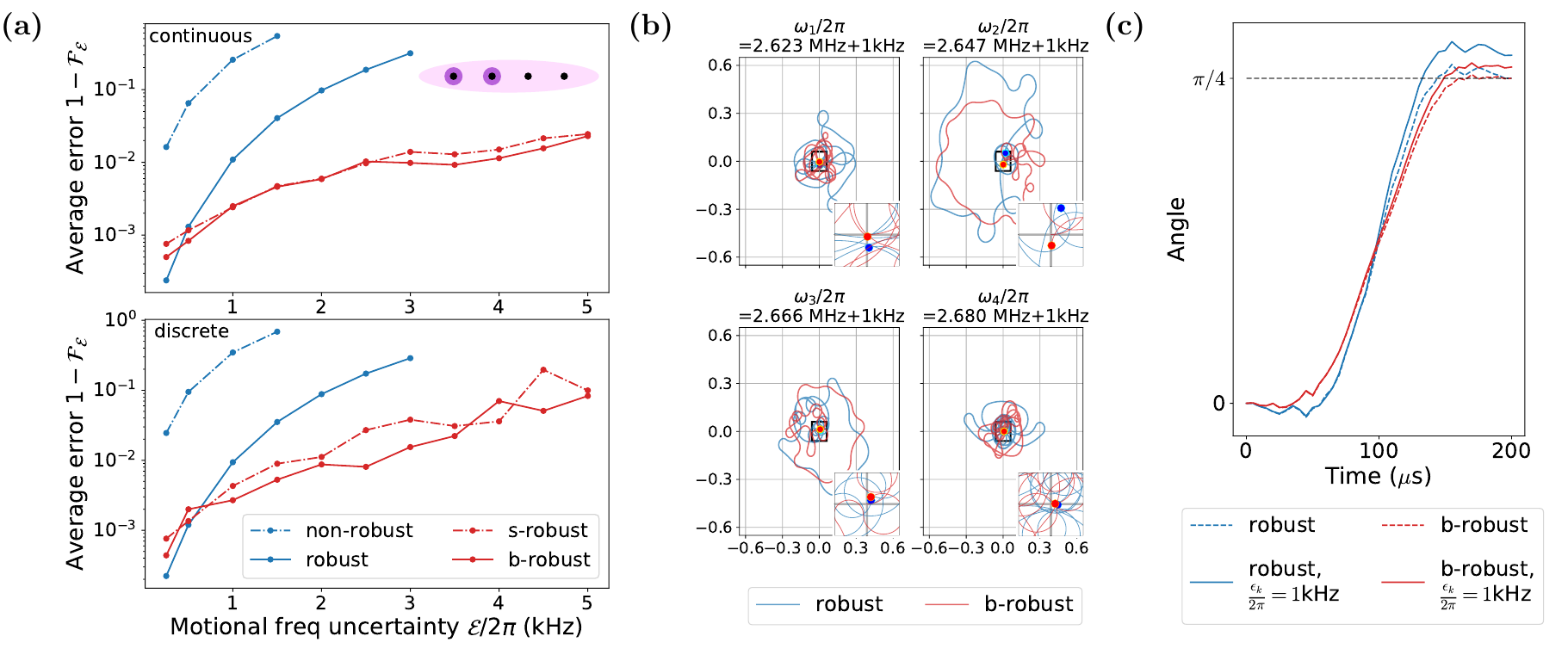}
 \caption{(a) Simulated unitary gate errors averaged over a test set of offsets drawn from distributions of various uncertainties $\mathcal{E}$. A 200-$\mu$s pulse on the first two ions of a four-ion chain is used. Each s- and b-robust pulse is optimized over the corresponding uncertainty $\mathcal{E}$. Except when $\mathcal{E}$ is too small, s- and b-robust FM are significantly more robust than robust FM. (b) Displacements (arbitrary units) of motional modes during the time when discrete robust and b-robust pulses are applied, where the mode frequencies drift by 1 kHz. The displacements at the end (circles) are overall closer to the origin when a b-robust pulse is applied. (c) Angle $\Theta(t)$ during the time when discrete robust and b-robust pulses are applied. When no drifts occur (dashed lines), the angle reaches exactly $\pi/4$ at the end of both the robust and the b-robust pulses. When a uniform drift of 1 kHz occurs (solid lines), the angle is closer to $\pi/4$ when a b-robust pulse is applied. For both (b) and (c), the b-robust pulse is optimized over \mbox{$\mathcal{E} = 2\pi \times 1$ kHz.}
 }
 \label{fig:2}
\end{figure*}
%%%%%%%%%%%END OF FIGURE 2%%%%%%%%%%%%%%%%

%%%%%%%%%%%%%%%%FIGURE 3 %%%%%%%%%%%%%%%%
\begin{figure*}[ht!]
\includegraphics[height=7.5cm]{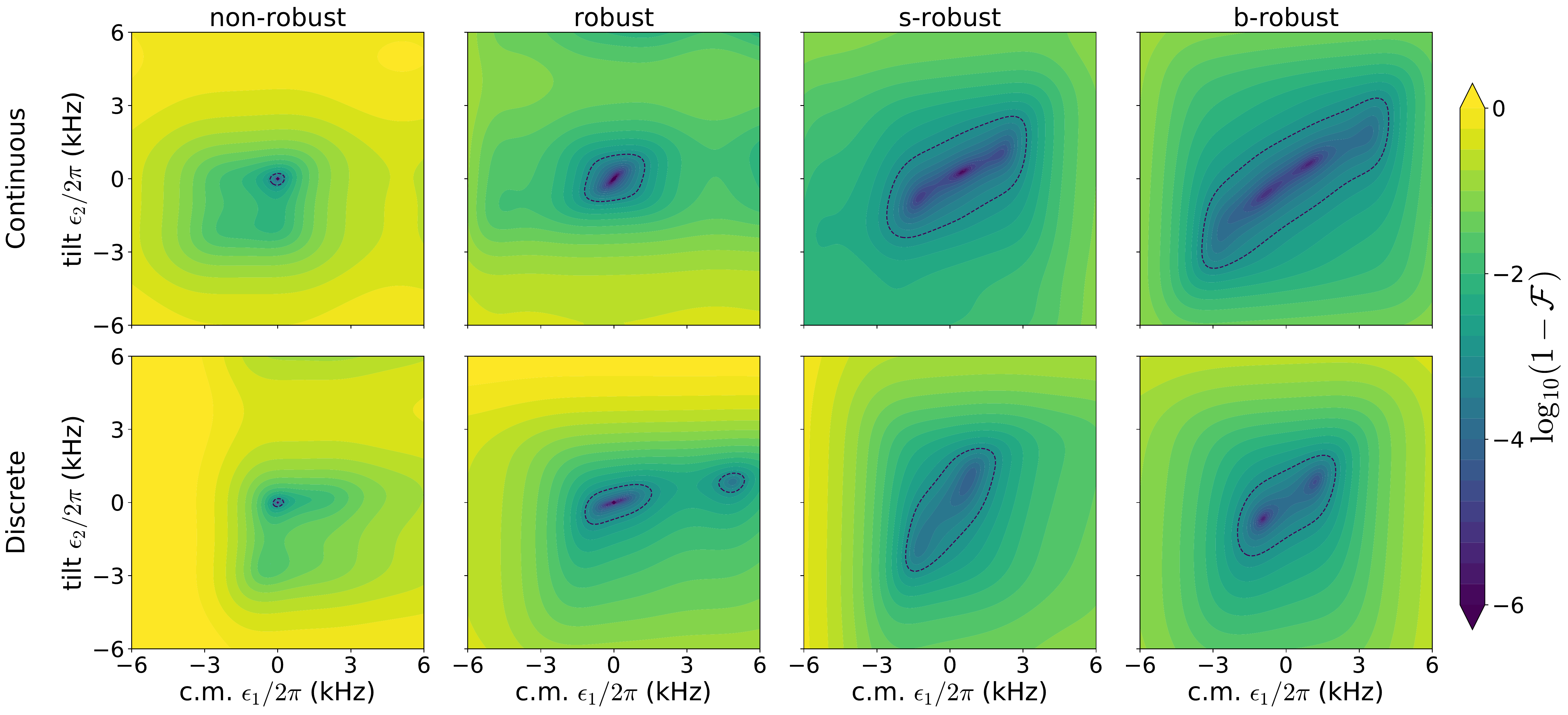}
\caption{Error landscapes over frequency offsets of motional center-of-mass (c.m.) and tilt modes, simulated for various FM pulses. A 200-$\mu$s pulse on a two-ion chain is used. The s- and b-robust pulses are optimized over a mode-frequency uncertainty $\mathcal{E} = 2\pi \times 1$ kHz. The regions where the error is lower than $10^{-3}$ are marked with dashed contour lines. The s- and b-robust pulses are clearly more robust over a wider region of offsets.}
\label{fig:3}
\end{figure*}
%%%%%%%%%%%END OF FIGURE 3 %%%%%%%%%%%%%%%%

Fig. \ref{fig:2}a shows the simulated average error \mbox{$1-\mathcal{F}_{\mathcal{E}}$} for pulses optimized by nonrobust, robust, s-robust, and b-robust FM, for various values of the mode-frequency uncertainty $\mathcal{E}$. We use a pulse of length \mbox{200 $\mu$s} to perform a MS gate on the first two ions in a four-ion chain. Note that each point of the s-robust and b-robust data is optimized with the respective range $\mathcal{E}$. We perform 300 iterations for nonrobust and robust FM and 1500 iterations for s- and b-robust FM. Since the optimization performance depends slightly on the choice of the random initial guess pulse, we perform ten trials and choose the optimized pulse with the best average fidelity over a cross-validation set, constructed randomly and independently from the test set. 

We find that s-robust and b-robust pulses have significantly smaller average errors than robust pulses, for an error range \mbox{$\mathcal{E}/2\pi \geq 0.5$ kHz} using continuous pulses, and for \mbox{$\mathcal{E}/2\pi \geq 1$ kHz} using discrete pulses. Notably, continuous s- and b-robust pulses have an average fidelity of approximately 0.99 over an offset range of \mbox{$\mathcal{E}/2\pi = 5$ kHz.} This shows that s- and b-robust FM can be robust to offsets as large as the inverse of the pulse length \mbox{$1/\tau = 5$ kHz.} In general, b-robust FM performs slightly better than s-robust FM, despite having ten times fewer cost-function and gradient evaluations than s-robust FM has. This can be understood as the advantage of exploring various values of offsets $\vec{\epsilon}$, thus reducing the gap between the training curve and the testing curve \cite{Wu-Ding19}.

Fig. \ref{fig:2}b and \ref{fig:2}c visualize the displacement and angle errors for robust and b-robust pulses where the motional mode frequencies drift by \mbox{$\epsilon_k/2\pi = 1$ kHz} for all four modes. As expected, the b-robust pulse has smaller errors in both displacement and angle. 

To visualize the robustness of the various FM methods, Fig. \ref{fig:3} plots the error landscapes over the motional frequency offsets. We use a pulse of length \mbox{200 $\mu$s} for a MS gate on a two-ion chain, with offsets of the center-of-mass mode ($\epsilon_1$) and tilt mode ($\epsilon_2$). Both continuous and discrete pulses are used. The s-robust and b-robust pulses are optimized over the mode-frequency uncertainty \mbox{$\mathcal{E}/2\pi = 1$ kHz.} The ``high-fidelity regions'' where the error is less than $10^{-3}$ are marked with dashed contour lines. For continuous pulses, the high-fidelity region is 4.5 and 6.4 times larger with the s-robust and b-robust pulses, respectively, than with the robust pulse. For discrete pulses, the high-fidelity region is 2.8 times larger with both the s-robust and the b-robust pulses than with the robust pulse. This shows that we can achieve significantly enhanced robustness with s-robust and b-robust FM. Also, note that the continuous b-robust pulse has a high-fidelity region 2.5 times larger than the discrete b-robust pulse has. 

The error landscapes for s-robust and b-robust pulses have two or three peaks of high fidelity that are clearly separated from the origin. The average position of the peaks is near the origin, thus guaranteeing high fidelity at zero offset as well. A large high-fidelity region that encompasses all peaks is formed. This provides an understanding of how s-robust and b-robust FM are able to achieve significantly better robustness than can robust FM, whose landscape has a single sharp peak at the origin. Note that a double-peak landscape is also observed in Ref. \cite{Wu-Ding19}, where minibatch optimization was performed over errors in coupling strengths. 

Now we discuss the scalability of robust and b-robust FM. Unlike various generic pulse-optimization algorithms whose computational cost increases exponentially with the number of qubits \cite{Grape, Krotov, Crab}, robust FM for a trapped-ion system has a linear computational cost, which makes the algorithm applicable to large-scale systems \cite{Leung18-E}. Our b-robust FM method also inherits this advantage. 

Fig.4 shows the performance of robust and b-robust FM optimized for ion chains in a harmonic trap potential, with the number of ions ranging from 2 to 12. For two- and four-ion chains, MS gates on all pairs of ions are simulated. For ion chains of length $N \geq 6$, MS gates on all pairs in a subchain of length $N-2$, excluding the ions at the edges, which are too weakly coupled to the motional modes, are simulated. The error bars indicate the standard deviation over the ion pairs. We use 400-$\mu$s pulses, both continuous and discrete. The b-robust pulses are optimized over a motional frequency uncertainty \mbox{$\mathcal{E}/2\pi = 0.5$ kHz.} For continuous b-robust optimization, we minimize only the displacement error \mbox{$C(\vec{\epsilon}) = \sum_k \Big(\alpha^{j_1}_k(\tau, \vec{\epsilon})^2 + \alpha^{j_2}_k(\tau, \vec{\epsilon})^2\Big)$} instead of the entire error as in Eq. \ref{eq:C}, due to computational-time issues. 1500 iterations are performed for each optimization. 

%%%%%%%%%%%%%%%%FIGURE 4 %%%%%%%%%%%%%%%%
\begin{figure}[ht]
 \includegraphics[width=8.6cm]{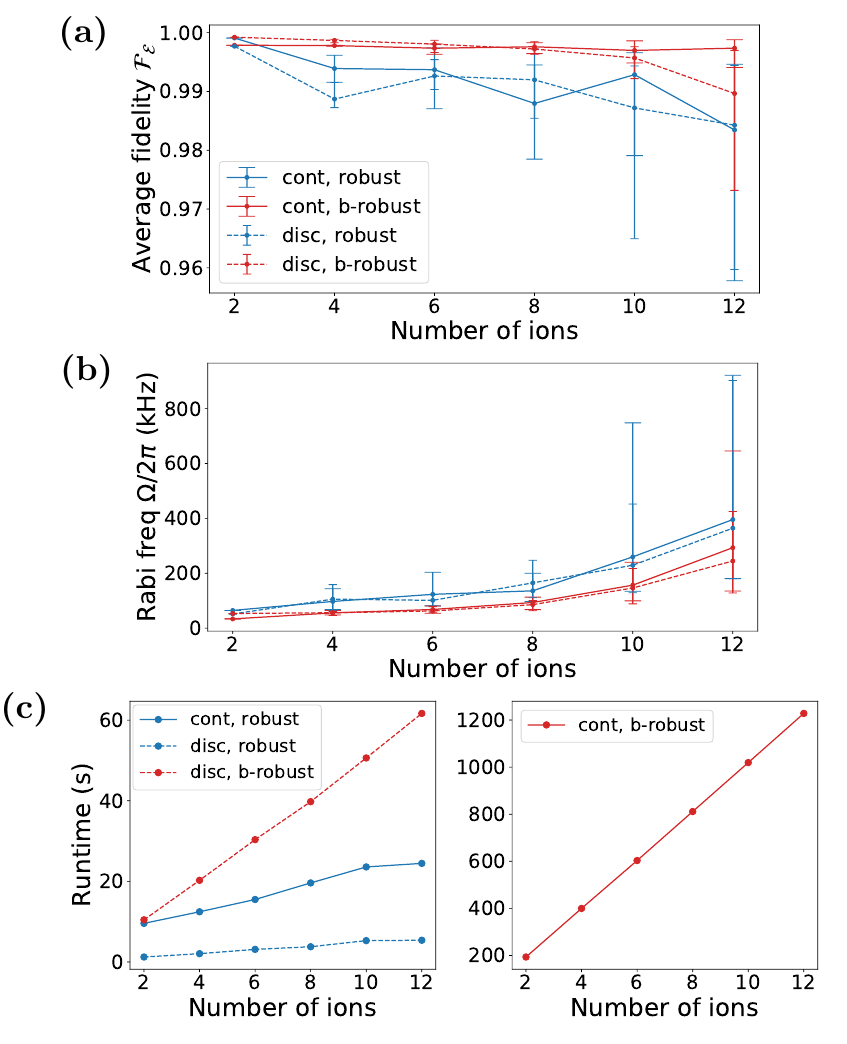}
 \caption{Scalability of robust and b-robust FM, simulated for up to a 12-ion chain. The error bars represent the standard deviation over all ion pairs on which gates are applied. For numbers of ions greater than or equal to 6, the ions at the edges are not used for gates. Both continuous and discrete pulses of length 400 $\mu$s are used. (a) Average fidelity over $\mathcal{E} = 2\pi \times 0.5$ kHz. (b) Rabi frequency. (c) Run time for single pulse optimization. Calculations are performed on a consumer laptop with a 1.60-GHz Intel Core i5 CPU and 16.0 GB RAM. }
 \label{fig:4}
\end{figure}
%%%%%%%%%%%END OF FIGURE 4 %%%%%%%%%%%%%%%%

Fig. \ref{fig:4}a plots the average fidelity $\mathcal{F}_\mathcal{E}$, where \mbox{$\mathcal{E}/2\pi = 0.5$ kHz,} and Fig. \ref{fig:4}b plots the Rabi frequency $\Omega$. For both continuous and discrete pulses, b-robust FM finds a pulse solution with higher average fidelity and lower $\Omega$, and the advantages become more significant as the number of ions increases. Note that explicit regularization of $\Omega$ is possible for both methods, but at the cost of lower average fidelity. Nonetheless, the cost function of b-robust FM for discrete (continuous) pulses scales as $\Omega^4$ ($\Omega^2$) with the frequency offset, which naturally leads to convergence to a low-$\Omega$ solution. We expect a further reduction in $\Omega$ can be obtained by carefully choosing the initial guess pulse for each pair of ions, as well as by designing the shape of the trap potential for an even spacing between ions \cite{Leung18-E}.

Fig. \ref{fig:4}c plots the run time for single pulse optimization with each FM method, executed on a standard consumer laptop. As expected, the run times scale linearly with the number of ions. The run time for discrete b-robust FM is more than ten times longer than that for discrete robust FM, due to the batch size of 10 and the additional computation of $\Theta(\tau)$. Nonetheless, even for a 12-ion chain, discrete b-robust FM optimizes within approximately 1 min, making it a practical candidate for actual experiments. 

For continuous b-robust FM, the run time is approximately 20 times longer than for discrete b-robust FM, even though we minimize only the displacement error. The most time-consuming routine is evaluating $\Omega \propto \Theta(\tau, \vec{0})^{1/2}$ and its gradient at each iteration, which is quadratic in the number of substeps in the continuous case. However, we still find continuous b-robust FM a promising scheme for larger-scale systems, as for a 12-ion chain, $\mathcal{F}_\mathcal{E}$ is significantly higher (average 99.7\% over ion pairs) than for the other FM methods. We note that the run times could be improved by parallelization using graphics processing units and the development of faster algorithms for continuous pulses.

\section{Experiment} \label{sec:experiment}

We compare experimental results for implementing discrete robust and b-robust FM pulses of length 120 $\mu$s on a two-ion chain of $^{171}$Yb$^+$. The detailed experimental setup is described in Ref. \cite{Wang20}. The rf source for modulating the control lasers is upgraded from direct digital synthesizers (AD9912) to a rf system-on-chip (ZCU111) driven by firmware from Sandia National Laboratories \cite{Clark2021}. 

FM and amplitude-modulation pulses require careful tracking of ac Stark shifts during the modulation sequence. This detail is suppressed in most derivations, since from a theoretical viewpoint it is simply bookkeeping. For our system, the dominant ac Stark shift is fourth-order by design, and FM leads to negligible changes in the Stark shift. This may not be the case for other ion qubits, where tracking the ac Stark shift will be critical. 

\begin{figure}[ht]
  \includegraphics[width=8.6cm]{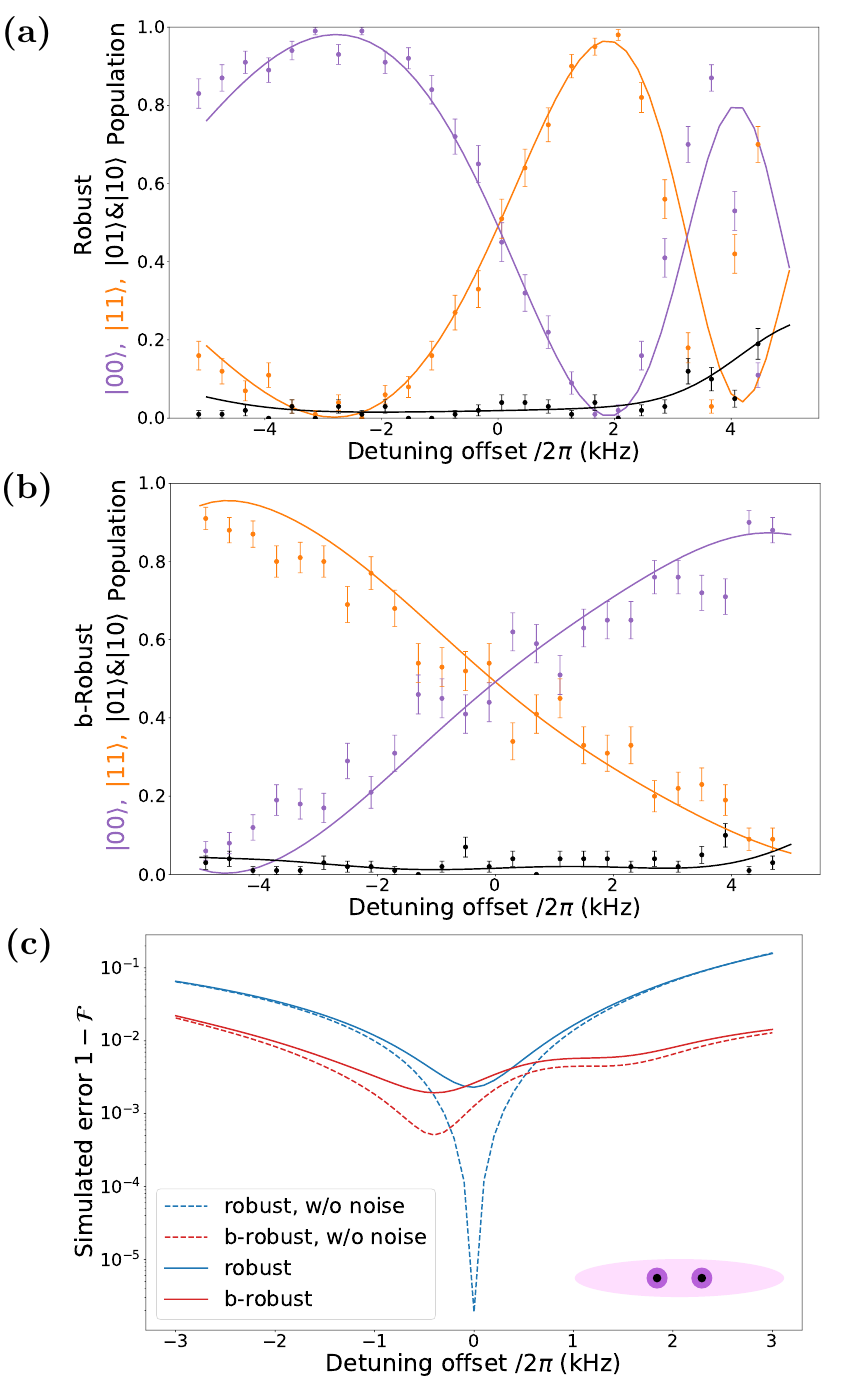}
 \caption{(a,b) Experimental (points) and simulated (lines) state populations over a range of detuning offsets, after sequences of five discrete (a) robust and (b) b-robust FM pulses are applied. The error bars represent the shot noise. The smaller slope of the even-parity curves and the flatter odd-parity curve indicate that b-robust FM is more robust than robust FM to detuning errors. (c) Gate errors averaged over sequences of five gates, simulated with (solid) and without (dashed) dissipative noise. The peak error is lower for b-robust FM in the presence of noise.}
 \label{fig:5}
\end{figure}

After initializing the qubits to the $\ket{00}$ state, we apply a sequence of five MS gates, which ideally generates the maximally entangled state $(\ket{00}+i\ket{11})/\sqrt{2}$.  To evaluate the effect of motional frequency drifts, we apply pulses with various detuning offsets. Fig. \ref{fig:5}a, \ref{fig:5}b show that with b-robust FM, the $\ket{00}$ and $\ket{11}$ populations deviate from 0.5 more slowly as the detuning offset increases, compared with robust FM. Also, the population of unwanted odd-parity states is more suppressed with b-robust pulses. This indicates that b-robust FM is more robust than robust FM to detuning errors. 

%%%%%%%%%%%%%%%%FIGURE 5 %%%%%%%%%%%%%%%%
%%%%%%%%%%%END OF FIGURE 5 %%%%%%%%%%%%%%%%

Fig. \ref{fig:5}c plots the simulated MS-gate errors for discrete robust and b-robust FM, both with and without dissipative noise. Each error is averaged over a sequence of five gates. We use a master equation \cite{Gardiner04} to simulate a MS gate under dissipative noise, which consists of motional dephasing, laser dephasing, and motional heating (see the Supplemental Material of Ref. \cite{Wang20} for details). The noise parameters that describe the current experiment are the following: motional coherence time 8 ms, laser coherence time 333 ms, and motional heating rate 400 and 40 quanta/s for the center-of-mass mode and the tilt mode, respectively. Although the peak gate fidelity for b-robust FM is lower than for robust FM without dissipative noise, it is slightly higher (99.81\%) than for robust FM (99.77\%) with noise. Appendix \ref{app:1} shows that b-robust FM is more robust than robust FM to slow dephasing noise, in the presence of motional frequency drifts. 

The peak fidelity for b-robust FM occurs at a detuning offset of $-0.4$ kHz. This is because optimizing over minibatches does not necessarily set the peak fidelity so that it is exactly at zero detuning. The simulations with dissipative noise predict that b-robust FM has slightly lower fidelity at zero detuning (99.74\%) than robust FM has. 

Appendix \ref{app:2} shows that the b-robust FM pulse achieves a MS-gate fidelity of 99.08(7)\% in the experiment. Note that this is lower than the MS-gate fidelity of 99.49(7)\% reported in Ref. \cite{Wang20}, where a robust FM pulse was used on the same system. The gate operates at zero detuning, calibrated to the point where a crossover between the populations of the $\ket{00}$ and $\ket{11}$ states occurs in the experiment described in Fig. \ref{fig:5}b. In future experiments with b-robust FM, the detuning offset should be calibrated to the expected gate-fidelity peak. Also, the gate suffers from a high heating rate of the transverse center-of-mass mode and off-resonant coupling to the motional modes in other directions, which is ignored in the gate-pulse design. We expect the gate fidelity to be improved when the trap is operated at a higher rf voltage, which corresponds to a higher transverse-mode frequency, lower heating rate, and smaller off-resonant coupling. However, the rf voltage in our experiments is currently limited by several malfunctioning electrodes in the surface trap. Although we disable those electrodes, the impedance of the trap changes, and the ions are observed to be unstable when the center-of-mass transverse-mode frequency is higher than $2\pi\times 2.1$ MHz.

\section{Conclusion}\label{sec:conclusion}

In this paper, we present s-robust and b-robust FM pulse optimization schemes for two-qubit entangling gates in trapped-ion systems. We improve on the robust FM scheme \cite{Leung18-R} by application of ML-inspired techniques, using a large sample set and minibatches, respectively. In our schemes, robustness is directly enforced by defining the cost function as displacement and angle errors averaged over various values of motional frequency offsets. Our results show that s- and b-robust FM achieve a robustness significantly improved from that of robust FM, finding pulse solutions with multiple peaks in the fidelity landscape. Scalability of b-robust FM, in terms of a high average fidelity, low laser-power requirement, and reasonable optimization run time, is demonstrated for up to 12 ions. Finally, we provide proof-of-concept experimental results that demonstrate improved robustness when using b-robust FM is used. We expect that b-robust FM has more significant advantages over robust FM in more complicated experiments with larger numbers of ions and uncertainty in the motional frequencies, as shown in Fig. \ref{fig:4}a and \ref{fig:2}a. 

Immediate directions include analyzing the trade-off of using the approximate error model in Eq. \ref{eq:U}-\ref{eq:C} versus using a more realistic model such as a master equation for the cost function of b-robust optimization. Another approach is to collect samples of gate errors at various parameter offsets directly with the experimental apparatus. 

The idea of b-robust FM can be extended to other types of pulse modulation and noise. One future direction is to extend the minibatch optimization scheme to find pulse solutions that are robust to fast time-varying noise when combined with quantum oscillator noise spectroscopy \cite{Milne20-Q}. In general, we expect that ML-inspired pulse-optimization tools for robust quantum control will make a significant contribution to high-fidelity operations, not only in trapped-ion systems but also on various other quantum computing platforms \cite{Wu-Ding19}.  

\begin{acknowledgments}
 This work was supported by the Office of the Director of National Intelligence, Intelligence Advanced Research Projects Activity through ARO Contract W911NF-16-1-0082, the National Science Foundation Expeditions in Computing Award 1730104, the National Science Foundation STAQ Project Phy-181891, and the U.S. Department of Energy, Office of Advanced Scientific Computing Research QSCOUT program.
\end{acknowledgments}

\appendix
\section{ROBUSTNESS TO DEPHASING NOISE}\label{app:1}
Minimizing the time-averaged displacement \mbox{$|\alpha^j_{k,\text{avg}}| \propto \frac{1}{\tau}|\int^\tau_0 \int^t_0 e^{-i\theta_k(t')}dt' dt|$} as in robust FM achieves robustness not only to systematic frequency offsets but also to time-dependent fluctuations in the motional mode frequencies and the laser amplitude, as demonstrated in both simulations and experiments with phase modulation \cite{Milne20-P}. Here we prove that minimizing $|\alpha^j_{k,\text{avg}}|$ achieves robustness to slow dephasing noise. Then we provide simulation results that show that b-robust FM is more robust to dephasing noise than robust FM is in the presence of motional frequency drifts, despite only minimizing the final displacements $|\alpha^j_k(\tau, \vec{\epsilon})|$.

Consider a time-dependent phase fluctuation $\varphi(t)$ caused by motional and/or optical dephasing noise. We assume the fluctuation is small: \mbox{$|\varphi(t)| \ll 1$ ($0 \leq t \leq \tau$)}. We also assume the fluctuation is slow compared with the inverse gate time:
\begin{equation}
    \varphi(t) = \frac{1}{\sqrt{2\pi}} \int^\infty_{-\infty} \tilde{\varphi}(\omega) e^{i\omega t}d\omega \approx \frac{1}{\sqrt{2\pi}} \int^{\omega_c}_{-\omega_c} \tilde{\varphi}(\omega) e^{i\omega t}d\omega
\end{equation}
where $\tilde{\varphi}(\omega)$ is the Fourier transform of $\varphi(t)$ and \mbox{$\omega_c \ll 1/\tau$} is the cutoff frequency. We consider the case where the final displacement is set to zero when there is no dephasing noise. Replacing the phase $\theta(t)$ with \mbox{$\theta(t) + \varphi(t)$,} we evaluate the displacement as in the following:

\begin{align}
    \alpha^j_{k}(\tau) &\propto \int^\tau_0 e^{-i[\theta_k(t) + \varphi(t)]} dt
    \approx \int^\tau_0 e^{-i\theta_k(t)}[1-i\varphi(t)] dt \nonumber \\
    &= -i\int^\tau_0 e^{-i\theta_k(t)} \varphi(t) dt \nonumber \\
    &\approx \frac{-i}{\sqrt{2\pi}} \int^{\omega_c}_{-\omega_c} d\omega\: \tilde{\varphi}(\omega) e^{i\omega t} \int^\tau_0 dt\:e^{-i\theta_k(t)} \nonumber\\
    &\approx \frac{-i}{\sqrt{2\pi}} \int^{\omega_c}_{-\omega_c} d\omega\: \tilde{\varphi}(\omega) \int^\tau_0 dt\:e^{-i\theta_k(t)}(1+i\omega t) \nonumber\\
    &= \frac{1}{\sqrt{2\pi}} \int^{\omega_c}_{-\omega_c} d\omega\:\omega \tilde{\varphi}(\omega) \int^\tau_0 dt\: te^{-i\theta_k(t)} \nonumber\\
    &= -\frac{1}{\sqrt{2\pi}} \int^{\omega_c}_{-\omega_c} d\omega\:\omega \tilde{\varphi}(\omega) \int^\tau_0 dt \int^t_0 dt' e^{-i \theta_k(t')} \nonumber \\
    &\propto \alpha^j_{k,\text{avg}},
\end{align}
where we perform integration by parts in the second to last step. Therefore we conclude that \mbox{$\alpha^j_{k,\text{avg}} \approx 0$} achieves first-order robustness to slow dephasing noise. 

To evaluate the robustness to slow dephasing noise in the presence of motional frequency drifts, we compute the time-averaged displacements averaged over a test set of motional frequency uncertainty $\mathcal{E}$:
\begin{equation}\label{eq:C1}
    \begin{gathered}
    C^{\text{avg}}_{\mathcal{E}} =  \frac{1}{|T_{\mathcal{E}}|} \sum_{\vec{\epsilon} \in T_{\mathcal{E}}} C^{\text{avg}}(\vec{\epsilon})
    \\
    C^{\text{avg}}(\vec{\epsilon}) = 
    \sum_k \big(\alpha^{j_1}_{k,\text{avg}}(\vec{\epsilon})^2 + \alpha^{j_2}_{k,\text{avg}}(\vec{\epsilon})^2\big)
    \end{gathered}
\end{equation}

\begin{figure}[ht!]
 \includegraphics[width=8.6cm]{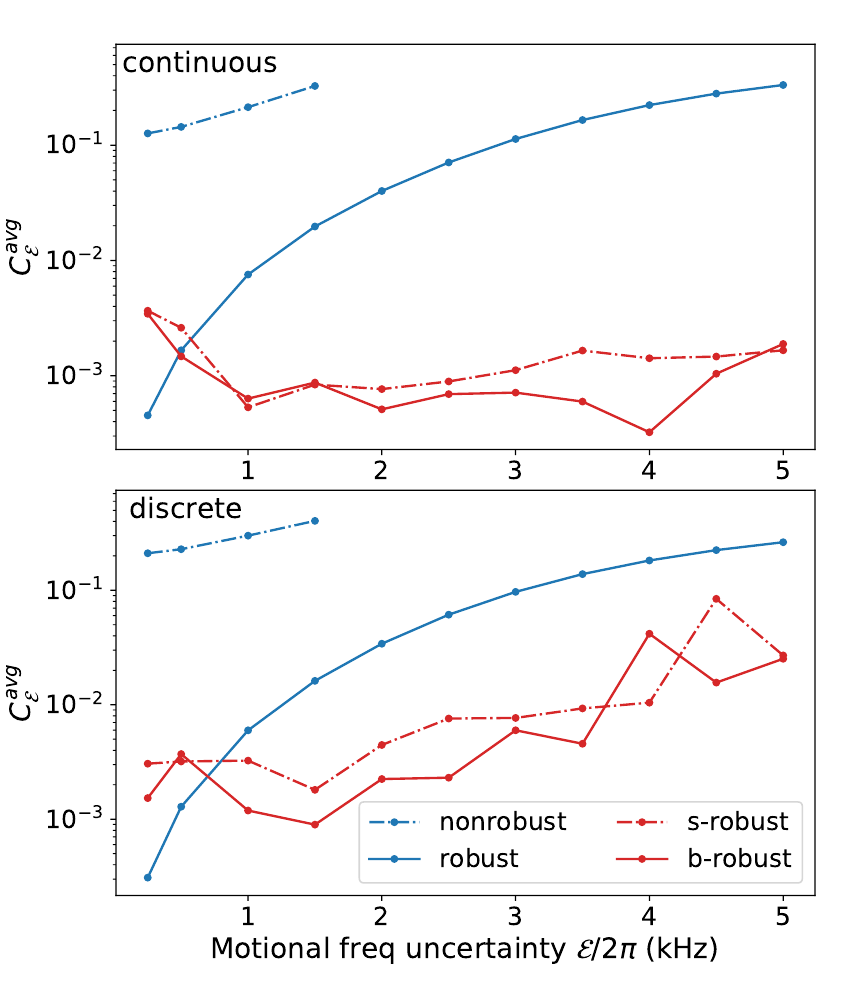}
 \caption{Time-averaged displacements averaged over a test set of offsets drawn from distributions of various uncertainties $\mathcal{E}$. A 200-$\mu$s pulse on the first two ions of a four-ion chain is used. Each s- and b-robust pulse is optimized over the corresponding uncertainty $\mathcal{E}$. Except when $\mathcal{E}$ is too small, s- and b-robust FM have significantly smaller time-averaged displacements than robust FM has. This implies that s- and b-robust FM are more robust to dephasing noise in the presence of motional frequency drifts.}
 \label{fig:6}
\end{figure}

Fig \ref{fig:6} plots $C^{\text{avg}}_{\mathcal{E}}$ for pulses optimized by nonrobust, robust, s-robust, and b-robust FM, for various uncertainties $\mathcal{E}$. As in Fig. \ref{fig:2}a, we use a 200-$\mu$s pulse on the first two ions in a four-ion chain. We find that s-robust and b-robust pulses have a significantly smaller $C^{\text{avg}}_{\mathcal{E}}$ than for robust FM when \mbox{$\mathcal{E}/2\pi \geq 1$ kHz.} While s- and b-robust FM minimize the final displacements over the uncertainty range, they naturally minimize the time-averaged displacements to satisfy the condition for robustness to motional frequency offsets. This also leads to robustness to dephasing noise, which shares the same condition. 

Notably, for continuous b-robust FM, \mbox{$C^{\text{avg}}_{\mathcal{E}} < 10^{-3}$} when \mbox{$1~\text{kHz} \leq \mathcal{E}/2\pi \leq 4~\text{kHz}$.} This leads to the displacement errors being reduced by 1-2 orders of magnitude compared with robust FM, because the displacement errors are proportional to the time-averaged displacements. In this range, the rotation-angle errors of the b-robust pulse dominate the displacement errors.

\section{EXPERIMENTAL GATE-FIDELITY MEASUREMENT}\label{app:2}

\begin{figure}[ht!]
 \includegraphics[width=8.6cm]{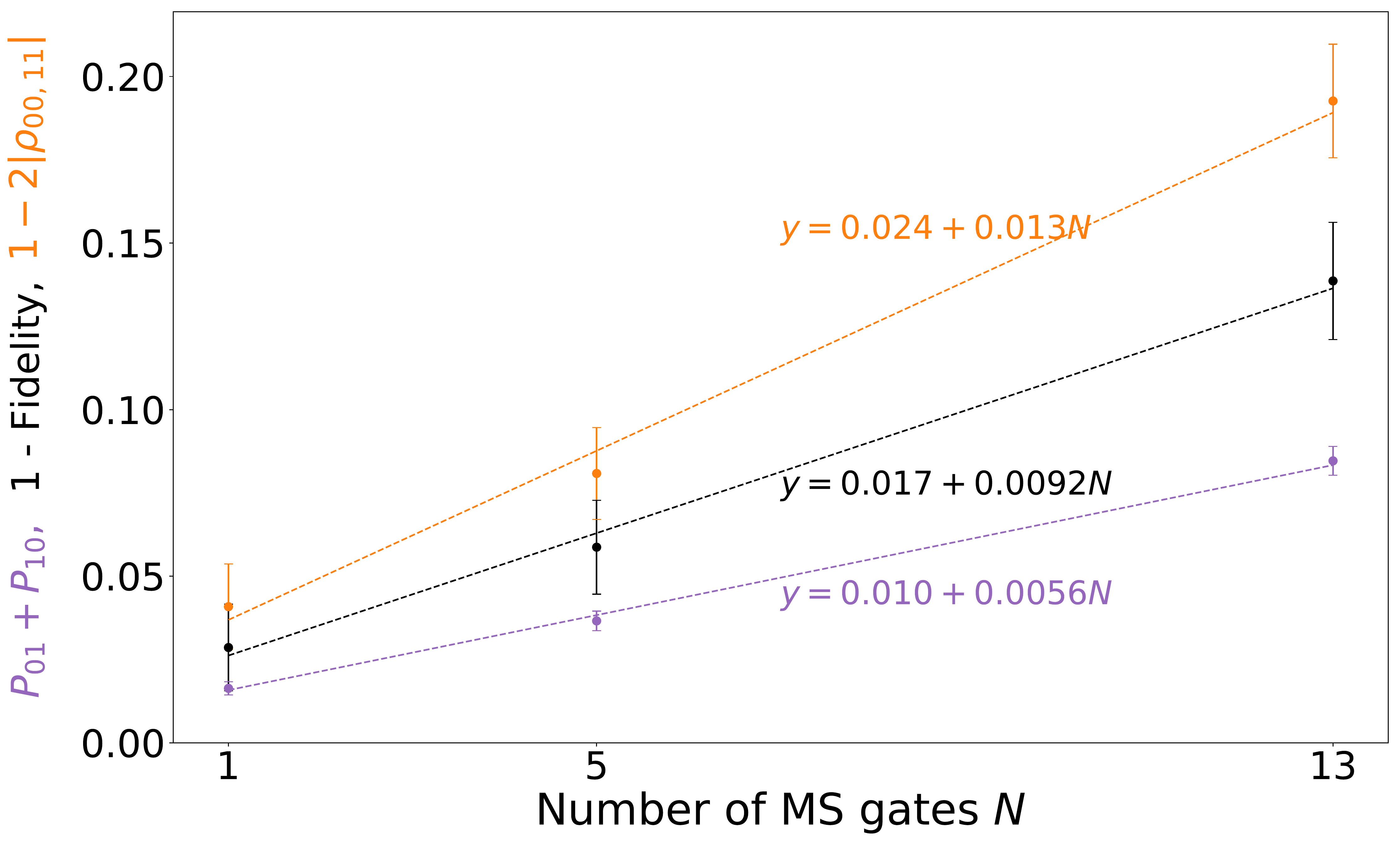}
 \caption{Experimental errors in the maximally entangled state generated by sequences of repeated MS gates. The purple, orange, and black points represent the population leakage to the $\ket{01}$ and $\ket{10}$ states, the loss of parity contrast, and the final-state error, respectively. The gate error is given by the slope of the linear fit to the black points.}
 \label{fig:7}
\end{figure}

We experimentally measure the MS-gate fidelity for discrete b-robust FM on a two-ion chain, using the method of Ref. \cite{Wang20}. We initialize the qubits to $\ket{00}$ and then apply a sequence of 1, 5, and 13 MS gates to generate the maximally entangled state \mbox{$(\ket{00} + i\ket{11})/\sqrt{2}$.} The population of the $\ket{01}$ and $\ket{10}$ states and the parity contrast are used to measure the state fidelity \cite{Leibfried03}. Using the fact that the stochastic error accumulates linearly, the coherent error accumulates quadratically, and the state-preparation-and-measurement (SPAM) error remains constant, we extract the gate fidelity without the SPAM error from a linear fit. From Fig. \ref{fig:7}, we measure the gate fidelity to be 99.08(7)\%. The data agree with the linear fit, indicating negligible coherent error.

\section{CONNECTIVITY OF A TEN-ION CHAIN}\label{app:3}

\begin{figure}[ht!]
 \includegraphics[width=8.6cm]{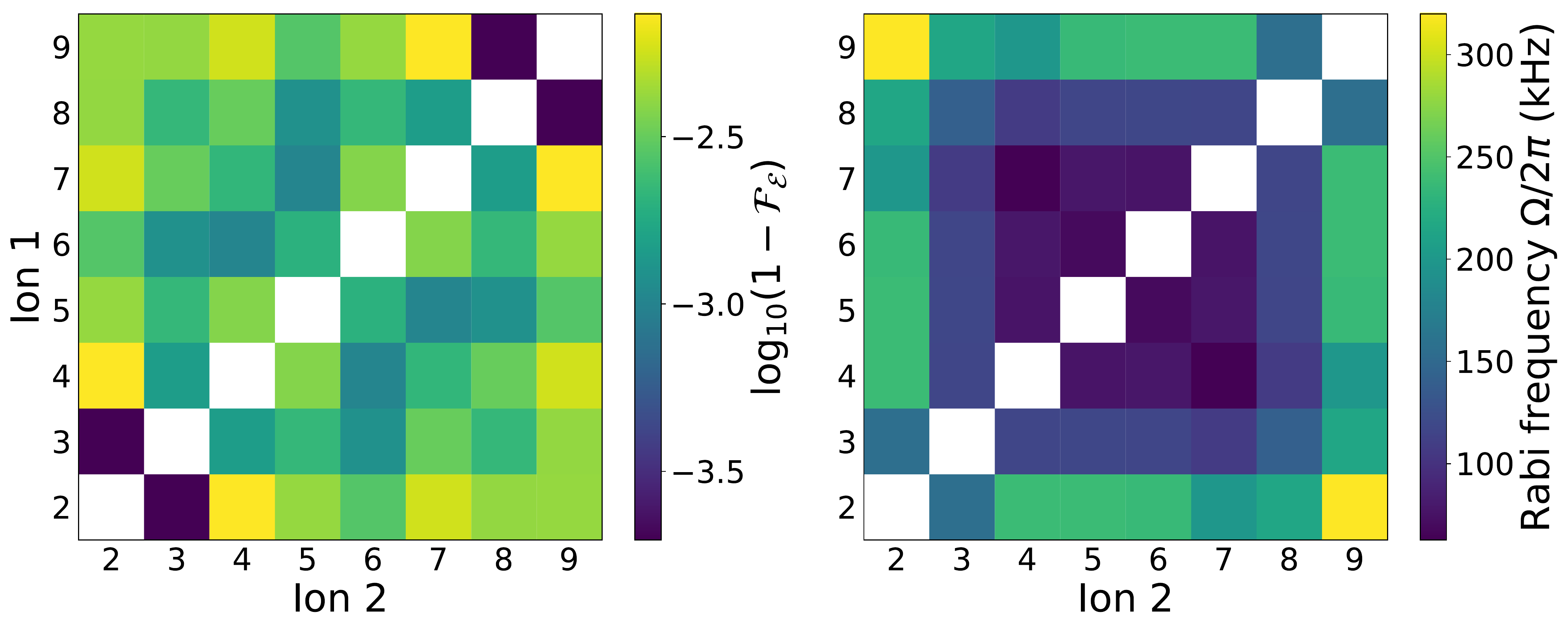}
 \caption{Connectivity of a ten-ion chain, simulated for continuous b-robust FM. The pulse length is 400 $\mu$s. Left: average error over motional frequency uncertainty $\mathcal{E} = 2\pi \times 0.5$ kHz. Right: Rabi frequency.}
 \label{fig:8}
\end{figure}

To understand the performance of b-robust FM in a larger system, Fig. \ref{fig:8} plots the connectivity of a ten-ion chain. A MS gate for each ion pair is optimized with continuous b-robust FM with a pulse length of 400 $\mu$s. The ions at the edges (1 and 10) are not used.

We expect to have a fully connected eight-qubit device with fidelities ranging from 0.993 to 0.9998, even with an uncertainty \mbox{$\mathcal{E} = 2\pi \times 0.5$ kHz} in the motional mode frequencies. A larger Rabi frequency is required for pairs that include the ion(s) close to the edges (2 or 9), due to smaller participation in the excited modes. This can be improved by carefully choosing the frequency offset of the initial guess pulse and shaping the trap potential to obtain evenly spaced ions \cite{Leung18-E}.

\section{BATCH SIZE AND OPTIMIZATION PERFORMANCE}

\begin{figure}[ht!]
 \includegraphics[width=8.6cm]{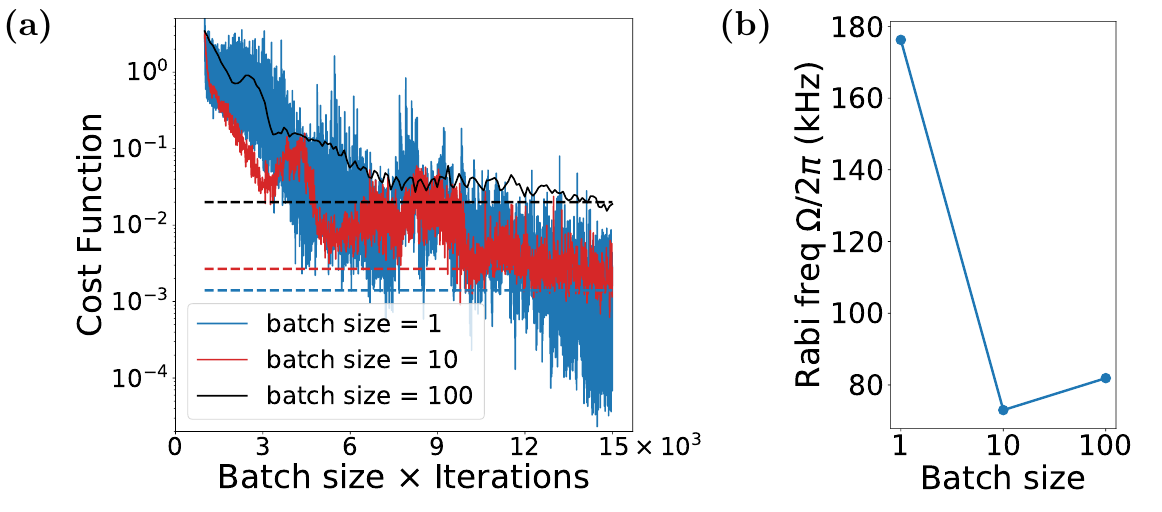}
 \caption{(a) Learning curves for discrete b-robust FM optimization on a four-ion chain with various batch sizes. The dashed lines represent the final average error $1-\mathcal{F}_{\mathcal{E}}$, where $\mathcal{E} = 2\pi \times 1$ kHz. (b) Rabi frequency of the optimized pulse for each batch size.}
 \label{fig:9}
\end{figure}

To describe the choice of the optimal batch size for b-robust optimization, Fig. \ref{fig:9} plots the learning curves for various batch sizes as in Ref. \cite{Wu-Ding19}, as well as the Rabi frequencies of the optimized pulses. A discrete pulse of length \mbox{200 $\mu$s} on the first two ions of a four-ion chain is used to optimize over the motional frequency uncertainty $\mathcal{E} = 2\pi \times 1$ kHz. Since the run time is proportional to the number of evaluations of the cost function $C(\vec{\epsilon})$ at a certain frequency offset, we fix the batch size times the number of iterations at 15000. 

Despite evaluating the cost function on the same number of samples, b-robust optimization with a smaller batch size leads to a higher fidelity $\mathcal{F}_\mathcal{E}$. This can be interpreted as the effect of batch-induced noise, which is shown in the fluctuations of the learning curve, leading to enhanced robustness \cite{Wu-Ding19}. However, when the batch size is 1, the Rabi frequency $\Omega$ of the optimized pulse is significantly higher than when the batch size is 10 or 100. This justifies our choice of the batch size as 10. We note that after parallelization, increasing the batch size while fixing the number of iterations does not necessarily increase the run time.

\bibliography{bib}% Produces the bibliography via BibTeX.

\end{document}